# Generating and Weaving Topological Event Wavepackets in Photonic Spacetime Crystals with Fully Energy-Momentum Gapped

Liang Zhang[1], Zirui Zhao[1], Qiaofei Pan[2], Chenhao Pan[1], Qingqing Cheng[3], Yiming Pan[1*]

[1] *State Key Laboratory of Quantum Functional Materials, School of Physical Science and Technology and Center for Transformative Science, ShanghaiTech University, Shanghai 200031, China*

[2] *Institute of Precision Optical Engineering, School of Physics Science and Engineering, Tongji University, Shanghai, 200092, China*

[3] *Shanghai Key Laboratory of Modern Optical System, School of Optical-Electrical and Computer Engineering, University of Shanghai for Science and Technology, Shanghai 200093, China.*

**Abstract**

We propose a novel type of topological excitation—topological event wavepackets (TEWs)—emerging in photonic spacetime crystals (STCs) with spacetime-modulated dielectric constants. These TEWs exhibit strong spatiotemporal localization and are topologically protected by a fully opened energy-momentum ($\omega k$) gap, within which conventional steady states are absent. We further demonstrate that TEWs are spectrally confined within the $\omega k$-gap, providing a combined measurement for probing the emergence of TEW and the $\omega k$-gap size. Furthermore, we construct a spacetime winding number to elucidate the protection of these events. Unlike previously reported nolinearity-induced event solitons, TEWs originate from topological configuration for linear media, thereby more accessible and versatile for experimental realization. Moreover, we show that TEWs can be periodically woven to form an event lattice, enabling to suppress unwanted noise amplification. Our findings open a new pathway toward topological control in photonic spacetime-modulated systems, enabling the $\omega k$-gap band engineering for wave manipulation ranging from microwave to optical regimes.

Research on time-varying media traces back to 1958, when Morgenthaler investigated how temporal variations in refractive index influence electromagnetic wave propagation [1]. For decades, progress in this area remained predominantly theoretical [2–6], with experimental efforts limited to microwave circuits exhibiting time-modulated behavior [7–9]. The modern concept of photonic time crystals (PTCs) was introduced around 2007 by Biancalana *et al.* [10] and 2016 by Shalaev *et al.* [11], catalyzed by advances in spatiotemporal dielectric structures and engineered metamaterials [12–19]. Notably, the first experimental realization of periodic temporal bandgaps—momentum gaps ($k$-gaps)—was demonstrated in a transmission line system by Halevi [20,21]. Building on this, Tretyakov *et al.* showed that a carefully designed microwave metasurfaces can support $k$-gaps shared by both surface-bound and free-space propagating modes [22], establishing a unified platform for wave control in time-varying environments.

Among the various architectures developed, photonic spacetime crystals (STCs) [see Fig. 1c]—structures with dielectric permittivities modulated periodically in both space and time—have emerged as a powerful paradigm for manipulating light beyond the limits of static or purely time-modulated media [23–31]. A defining feature of STCs is the formation of complete energy-momentum ($\omega$-$k$) gaps in their dispersion relations [27,29], arising from the simultaneous breaking of temporal and spatial translational symmetries. This dual symmetry breaking invalidates the conservation of energy and momentum, permitting even infinitesimal perturbations (such as noise) within the gap to undergo exponential amplification, thereby posing a fundamental challenge to dynamical stability. In our prior work, we showed that such instabilities can self-organize into nonlinear localized modes known as event solitons [29]. However, a general linear mechanism for stabilizing $\omega k$-gapped systems and suppressing uncontrolled noise growth has remained an open question.

To address this challenges, topological photonics provides a powerful framework for engineering wave systems that are robust against disorder and imperfections [32–34]. In spatially periodic photonic crystals (SPCs), topological invariants such as the Zak phase—a Berry phase accumulated across a one-dimensional Brillouin zone—predict the emergence of localized states at interfaces between regions of distinct topological phase [35,36]. These ideas have been successfully extended to temporally periodic systems, including photonic time crystals (PTCs), where topological edge states arise at temporal kinks separating Floquet phases with distinct quasienergy band structures [37,38], and it has been confirmed experimentally [39]. A. Szameit *et al.* demonstrated experimentally a topological event by launching an off-shell excitation that experiences the separated momentum gap ($k$-gap) and frequency gap ($\omega$-gap) [40].

Collectively, these studies demonstrate that topological protection can persist in non-stationary, time-dependent systems. However, such approaches rely on well-defined Bloch-Floquet band structures formulated in either momentum ($k$) or frequency ($\omega$) space. Unlike purely spatially or remporally periodic systems, spatiotemporally modulated media exhibit bandgaps in the energy-momentum ($\omega$-$k$) space, which prevents the Brillouin zone from forming an integral loop. This breakdown of conventional band theory necessitates a spacetime-based topological framework capable of capturing topological features beyond Bloch paradigms.

Building on this perspective, we adopt a real-space topological approach inspired by the Jackiw–Rebbi model [41–44], which describes zero-energy modes bound to domain walls in Dirac systems. Qingqing Cheng *et al.* proposed and demonstrated a topologically protected plasmonic interface mode in a metallic ridge waveguide array [45]. Extending these concepts to (1+1)D spacetime, we engineer domain walls by modulating the refractive index in both space and time, thereby generating spatiotemporal topological kinks. These kinks support strongly localized wavepackets that are protected by a spacetime winding number $w$, circumventing the need for band theory. This perspective offers both conceptual clarity and practical control over topological event manipulation in STCs.

In this work, we introduce a new class of such spatiotemporal excitations in STCs, termed topological event wavepackets (TEWs). These events are sharply localized in both space and time, with their central frequency and momentum residing within a complete $\omega k$-gap. They emerge at spatiotemporal kinks and are protected by spacetime winding numbers. Notably, the momentum-spectral width of the TEW directly encodes the size of the $\omega k$-gap, offering a novel method for probing the gap profile through its localized dynamics. We further propose a spatiotemporal weaving strategy that forms a topological event lattice, which enables to suppress mode instabilities within the gap. Taken together, our results establish a comprehensive framework for the generation and deployment of topological events in photonic STC systems.

**Energy-momentum gap from Spatiotermporally Modulated Maxwell's Equations**. To understand the emergence of topological event wavepackets (TEWs) in Fig. 1f, we begin by modeling light propagation in a linear, isotropic, and non-magnetic dielectric medium with no free charges or currents (i.e., $\rho = 0, \boldsymbol{J} = 0$). Maxwell's equations in a source-free medium are given by $\nabla \times \boldsymbol{E} = -\frac{\partial \boldsymbol{B}}{\partial t}$, $\nabla \times \boldsymbol{B} = \mu_0 \frac{\partial \boldsymbol{D}}{\partial t}$, with divergence equations $\nabla \cdot \boldsymbol{D} = 0, \nabla \cdot \boldsymbol{B} = 0$, where **E**, **D** and **B** are the electric field, electric displacement field, and magnetic field, respectively, and $\mu_0$ is the vacuum permeability. Applying the curl operator to Faraday's law and substituting into

Ampère's law yields wave equation $\frac{\partial^2 \boldsymbol{D}}{\partial t^2} = -\frac{1}{\mu_0} \nabla \times (\nabla \times \boldsymbol{E})$.

Assuming a non-magnetic, linear dielectric response, we write $D = \epsilon(x,t)E = \epsilon_0 \epsilon_1(x,t)E$, where $\epsilon_1(x,t)$ represents the modulated dielectric function. Substituting into the generalized wave function gives $\frac{\partial^2 D}{\partial t^2} = \frac{1}{\mu_0} \frac{\partial^2}{\partial x^2} \left(\frac{D}{\epsilon_0 \epsilon_1}\right)$. In time-varying photonic media, D is often preferred due to its continuity across temporal interfaces, as required by Gauss's law [46]. In contrast, the electric field E can be discontinuous, giving rise to time reflection and refraction [6]. The linear dielectric constant, $\epsilon_1$, is modulated periodically in space and time, and we write it as $\epsilon_1(x,t) = \epsilon_r \tilde{\epsilon}(x)\tilde{\epsilon}(t) = \epsilon_r(1 + \delta_1 cos(\Omega t))(1 + \delta_2 cos(Gx))$. Here, $\epsilon_r$ denotes the mean relative permittivity, and $\delta_{1,2}$ represent the small modulation depths in time and space, typically $0.1 \sim 0.3 \, \epsilon_{eff}$ in experimental settings. The temporal and spatial modulation frequencies are given by $\Omega = 2\pi/T$, $G = 2\pi/\Lambda$, corresponding to temporal and spatial periods T and $\Lambda$. For an optical implementation using an 800 nm laser, a spatial period of $\Lambda = 400nm$ and temporal modulation period of $T = 1.3fs$ are suitable. We approximate $\tilde{\epsilon}^{-1}(t) = 1 - \delta_1 cos\Omega t$. The wave function can be simplified to $\frac{(1+\delta_2 cosGx)}{c^2} \frac{\partial^2 \tilde{E}}{\partial t^2} = (1 - \delta_1 cos\Omega t)\frac{\partial^2 \tilde{E}}{\partial x^2}$, where $\tilde{E} = D/\tilde{\epsilon}(x)$ is the reduced electric displacement, and $c = c_0/\sqrt{\epsilon_r} = c_0/n_0$ is the effective speed of light in the background medium, with $n_0 = \sqrt{\epsilon_r}$ being the refractive index and $c_0$ the light speed in vacuum. This modulated wave equation captures the dynamics of electromagnetic waves in a spatiotemporally varying medium, serving as the starting point for analyzing topological phenomena in STCs.

To derive the topological event wavepacket, we simplify it further using coupled-mode equations based on the Floquet-Bloch theorem. Assuming Floquet-Bloch waves as a sum of spatiotemporally modulated forward and backward waves, we write

$$\tilde{E}(x,t) = A_f e^{i\frac{G}{2}x - i\frac{\Omega}{2}t} + A_b e^{-i\frac{G}{2}x - i\frac{\Omega}{2}t} + A_f^* e^{-i\frac{G}{2}x + i\frac{\Omega}{2}t} + A_b^* e^{i\frac{G}{2}x + i\frac{\Omega}{2}t}, \quad (1)$$

where $A_f$, $A_b$ are the complex amplitudes of the forward- and backward-propagating waves, and their conjugates $A_f^*$, $A_b^*$. Applying the slowly-varying envelope approximation [47] ($\delta_{1,2} \ll 1$) yields the 4*4 Dirac-type equations:

$$iG\frac{\partial A_f}{\partial x} + i\frac{\Omega}{c_r^2}\frac{\partial A_f}{\partial t} + \left(\frac{\Omega^2}{4c_r^2} - \frac{G^2}{4}\right)A_f + \frac{\delta_2 \Omega^2}{8c_r^2}A_b + \frac{\delta_1 G^2}{8}A_b^* = 0$$

$$-iG\frac{\partial A_b}{\partial x} + i\frac{\Omega}{c_r^2}\frac{\partial A_b}{\partial t} + \left(\frac{\Omega^2}{4c_r^2} - \frac{G^2}{4}\right)A_b + \frac{\delta_2\Omega^2}{8c_r^2}A_f + \frac{\delta_1 G^2}{8}A_f^* = 0$$

$$-iG\frac{\partial A_f^*}{\partial x} - i\frac{\Omega}{c_r^2}\frac{\partial A_f^*}{\partial t} + \left(\frac{\Omega^2}{4c_r^2} - \frac{G^2}{4}\right)A_f^* + \frac{\delta_2\Omega^2}{8c_r^2}A_b^* + \frac{\delta_1 G^2}{8}A_b = 0$$

$$iG\frac{\partial A_b^*}{\partial x} - i\frac{\Omega}{c_r^2}\frac{\partial A_b^*}{\partial t} + \left(\frac{\Omega^2}{4c_r^2} - \frac{G^2}{4}\right)A_b^* + \frac{\delta_2\Omega^2}{8c_r^2}A_f^* + \frac{\delta_1 G^2}{8}A_f = 0. \quad (2)$$

These equations capture the scattering between Floquet-Bloch waves, including Bragg reflection ($A_f \leftrightarrow A_b$), and time reflection ($A_f \leftrightarrow A_b^*$). To simplify the above equations, we rewrite the Floquet-Bloch waves as a spinor $\psi = (A_f, A_b, A_f^*, A_b^*)^T$. And we define a dimensionless modulation ratio: $r = \Omega/Gc$, characterizing the balance between temporal and spatial modulations. Equation (2) can be recast in a compact Dirac-like form $\left(\left(\frac{i}{G}\frac{\partial}{\partial x}\right)\sigma_z\tau_z + \left(\frac{ir}{Gc}\frac{\partial}{\partial t}\right)\sigma_z\tau_0 + \frac{r^2-1}{4}\sigma_0\tau_0 + \frac{\delta_1}{8}\sigma_x\tau_x + \frac{\delta_2 r^2}{8}\sigma_0\tau_x\right)\psi = 0$, where $\sigma_i, \tau_j$ are Pauli operators, with $\sigma_i\tau_j = \sigma_i \otimes \tau_j$ for $i,j = 0, x, y, z$. To derive the band structure of the modulated photonic spacetime crystal, we consider plane-wave solutions of the form $\psi = \chi e^{iPx-iEt}$, where $\chi$ is a constant four-component spinor, $P$ is the effective wavevector and $E$ is the effective frequency. Substituting this ansatz into Eq. 2 yields,

$$\left(\frac{P^2}{G^2} - \frac{E^2 r^2}{c^2 G^2} - \frac{(r^2-1)^2}{16} - \frac{\delta_1^2}{64} + \frac{\delta_2^2 r^4}{64}\right)^2 = \frac{(r^2-1)^2}{4}\left(\frac{E^2 r^2}{c^2 G^2} + \frac{\delta_1^2}{64}\right). \quad (3)$$

This dispersion relation reveals the mode hybridization and band engineering induced by both temporal and spatial modulations [Fig. 1f]. In particular, when the modulation ratio $r \to 1$, forward/backward and time-reflection/refraction waves become strongly coupled, resulting in a fully opened $\omega k$-gap in the dispersion relation. Within this gap, four degenerate dissipative modes emerge, given by $\chi_1 e^{\eta t - \xi x}, \chi_2 e^{\eta t + \xi x}, \chi_3 e^{-\eta t + \xi x}$, and $\chi_4 e^{-\eta t - \xi x}$, where the spinor $\chi_m, m = 1, 2, 3, 4$ denote the components of the hybrid waves. These gapped modes represent all possible combinations of spatial growth (decay) and temporal amplification (attenuation), reflecting the rich modal landscape within the $\omega k$-gap.

**Topological Event Wavepackets.** To understand the topological hybridization of the unstable $\omega k$-gapped modes, we adopt a spacetime topological framework that circumvents the limitation of band theory. In spatiotemporally modulated systems, neither $k$ nor $\omega$ is conserved due to the simultaneous breaking of both translational symmetries [48]. To resolve this, we draw inspiration from the Jackiw-Rebbi solution in one-dimensional Dirac systems and extend this framework to the (1+1)D spacetime domain. Analogous to the Dirac equation, we interpret the modulation amplitudes $\delta_1$ and $\delta_2$ as effective mass terms, generalized to spatiotemporal profiles $\delta_1(t)$ and $\delta_2(x)$. Specifically, we introduce kink-like modulations for the effective mass terms $\delta_1(t) = \kappa_1 \tanh(10t), \delta_2(x) = \kappa_2 \tanh(10x)$, where $\kappa_1$ and $\kappa_2$ are the maximal strength of temporal and spatial modulations, respectively. Here, $\kappa_1$ governs the momentum gap, directly controlling its size, while $\kappa_2$ regulates the energy gap, with a similarly direct influence on the gap width. These profiles form a spacetime domain wall, at the intersection of which topologically localized event wavepackets emerge.

To facilitate analytical progress, we approximate the smooth domain-wall profiles $\tanh(10t)$ and $\tanh(10x)$ by their sharp counterparts—the sign functions: $\delta_1 \approx \kappa_1 sgn(t), \delta_2 \approx \kappa_2 sgn(x)$. This approximation preserves the essential topological character of the kink-like mass terms while substantially simplifying the mathematical treatment. The use of $sgn(x)$ is a standard approach in topological field theory, particularly in the original Jackiw–Rebbi model, where it enables closed-form bound-state solutions [42]. With this approximation, the modulation amplitudes in the four quadrants of the (x, t) plane can be considered approximately uniform, reducing the problem to a piecewise constant configuration. The domain walls at $x = 0$ and $t = 0$ intersect to form a cross-shaped spacetime topological interface, which acts as a trap for localized wavepackets. Following the Jackiw–Rebbi method, we solve the Dirac-type equation in each quadrant and match the wavefunction across boundaries to ensure physical admissibility. This yields a normalized analytical solution for the topological event wavepacket (TEW)

$$\psi(x,t) = \frac{1}{2}\begin{pmatrix} 1 \\ -i \\ 1 \\ i \end{pmatrix} e^{-|\kappa_1|\frac{Gct}{8} - |\kappa_2|\frac{\Omega x}{8c}}. \tag{4}$$

This solution exhibits exponential localization in both time and space, with decay rates controlled by the modulation strength $\kappa_1, \kappa_2$ and frequencies $\Omega, G$. Quantitatively, the corresponding root-mean-square (RMS) widths of the TEW are $\Delta t_{RMS} = 8/\sqrt{2}\kappa_1 Gc$, $\Delta x_{RMS} = 8c/\sqrt{2}\kappa_2\Omega$, indicating that stronger and faster modulations lead to tighter spatial and temporal confinement. This analytical result highlights the sensitivity of

TEWs to the underlying spatiotemporal modulation, offering a clear physical interpretation of their confinement. From a broader perspective, the TEW arises from a spacetime hybridization, where the four degenerate dissipative modes inside the $\omega k$-gap are coupled via spacetime kink engineering. Its confinement is protected by the topological mismatch across spacetime kinks, rendering the TEW stable against both perturbations. This makes TEWs not only a conceptually novel excitation, but also a promising experimental probe of the $\omega k$-gap topology.

**Protection of TEWs.** To further investigate the dynamic generation of topological event wavepackets (TEWs), we numerically simulate the Dirac-type equation with a localized Gaussian input propagating through a spacetime kink configuration. The effective Dirac mass terms are defined as $\delta_1 = \kappa_1 \tanh[10(t - 40T)]$ and $\delta_2 = \kappa_2 \tanh(10x)$, forming a spacetime kink centered at (x, t)=(0, 40T), where the weak seed is amplified to a sufficient strength. To characterize the topological nature of this configuration, we define a spacetime winding number $w$, based on the phase of the mass vector ($\boldsymbol{\delta} = (\delta_1, \delta_2)$):

$$w = \frac{1}{2\pi} \oint_c \nabla_{x,t} \theta(x,t) \cdot d\boldsymbol{l}, \tag{5}$$

where $\theta(x,t) = \tan^{-1}(\delta_1/\delta_2)$, and the contour $c$ encircles the domain wall in spacetime. For the kink profile above, we find $w = 1$, indicating a nontrivial spacetime topology. With modulation frequency $\Omega = G = 20\pi$ and modulation strengths $\kappa_1 = \kappa_2 = 0.3$, this relatively large modulation strength is chosen to clearly open the $\omega k$-gap, enhancing the visibility of topological phenomena. To initiate the system's dynamics, we launch a weak Gaussian wavepacket $0.0001\exp(-x^2/2\sigma^2)$ with $\sigma = 10$ at $t = 0$, and simulate its evolution under the Dirac-type equation. This setup enables us to observe how the input couples to the topological structure of the $\omega k$-gap and whether a topological event wavepacket emerges dynamically in the vicinity of the kink.

A sharply localized topological event wavepacket forms at the kink location, exhibiting strong confinement in both space and time [Fig. 2a]. To examine its sensitivity to structural gradients, we implement a non-uniform modulation profile, where the effective mass terms vary linearly in time and space $\delta_1(t) = -0.1(t - 40T)$, $\delta_2(x) = 0.1x$. The TEW persists, albeit with slight broadening[Fig. 2d]. This demonstrates that the event localization is shaped by the local slope and curvature of the kink, rather than exact symmetry, confirming the robustness of TEWs against modulation inhomogeneity. Importantly, the sign of the winding number (e.g., $w = -1$) does not affect the existence of TEWs [Figs. 2b, 2e]. These findings underscore the spactime topological protection of those events and their potential to persist in inhomogeneous, noise-amplifying media, indicating a viable route for localized energy confinement and signal control in dynamically unstable time-varying systems.

Fourier analysis of the TEW reveals that its energy and momentum spectra are entirely confined within the fully opened $\omega k$-gap, where no extended states exist [Fig. 2c]. This full confinement complements its spacetime localization, establishing the TEW as a genuine topological bound state. From the TEW envelope, the spectral root-mean-square (RMS) widths scale as $\Delta k = \kappa_2 \Omega/8c$, $\Delta \omega = \kappa_1 Gc/8$, which reflect the underlying modulation scales and are approximately half the widths of the momentum and energy bandgaps, respectively. To quantitatively validate this behavior, we systematically measured how $\Delta k$ and $\Delta \omega$ change with the modulation amplitudes $\kappa_1$ and $\kappa_2$, and compared the results with theoretical predictions. Both momentum and spectral widths exhibit linear dependence on the corresponding modulations, in excellent agreement with analytical estimates [Fig. 2f]. This connection indicates that TEW can serve as a sensitive probe for measuring the gap width, offering an approach for gap metrology in spacetime crystals, without requiring direct band-edge detections or transmission scans. More broadly, the TEW can be viewed as a classical analogue of a relativistic event—localized in both space and time, and characterized by a well-defined momentum $P_0$ and energy $E_0$, which are fully gapped by the modulations. It thus represents a topologically protected, transient excitation in spacetime continuum.

**Formation of Topological Event Lattices.** A hallmark of $\omega k$-gapped systems is their tendency to absorb energy and exponentially amplify fluctuations under spatiotemporal modulation. This behavior presents significant challenges for the physical realization of time-varying photonic systems. Our numerical simulations reveal that topological event wavepackets (TEWs), beyond acting as localized gap modes, can delay or suppress the onset of noise amplification, effectively stabilizing the system. To investigate this effect systematically, we construct a topological event lattice by periodically weaving spacetime kinks along both spatial and temporal axes. This "spatiotemporal weaving" effectively suppresses background growth, showing that domain wall arrangements play a key role in enhancing the system's dynamical stability within the $\omega k$-gap.

To quantitatively evaluate how spatiotemporal weaving of TEWs suppresses noise amplification, we fix the spatial domain periodicity at $X = 20\Lambda$, and systematically vary the temporal repetition cycle $T_R$ of the spacetime kinks. Exciting the system with a weak plane wave while keeping all other parameters unchanged. Figures 3a–e illustrate the resulting event lattices for $T_R = 3.8T, 4T, 10T, 18T$ and $25T$. Spatiotemporally localized TEWs are clearly observed in all cases. For short repetition periods ($T_R < 3.8T$), suppression of background amplification is most effective, but such rapid switching may pose experimental challenges. As $T_R$ increases, the suppression capability initially deteriorates, reaching a minimum near $T_R = 10T$. Remarkably, further increasing $T_R$ leads to improved suppression, peaking around $T_R = 18T$, near the full width at half maximum of the wave packet, and followed by a gradual decline at larger $T_R$ [Fig. 3f]. This non-monotonic dependence reveals the presence of an optimal temporal weaving period, suggesting that topological event lattices can be engineered to maximize dynamical stability within the $\omega k$-gap.

**Further discussions.** From an experimental standpoint, the realization of TEWs is feasible in both the microwave and optical regimes, albeit with distinct technical requirements. In the microwave domain, spacetime modulation can be implemented via varactor-loaded transmission lines [21], programmable metasurfaces [22] or active circuit networks [49] with tunable dielectric properties. Temporal modulation frequencies in the MHz to GHz range and spatial periods on the order of centimeters, are readily accessible with current RF electronics [50] and programmable microwave components. In contrast, the optical regime presents more stringent challenges, as it requires ultrafast temporal modulation on the femtosecond timescale—for instance, ~1.3 fs for an 800 nm carrier. However, recent advances in epsilon near zero (ENZ) materials [51] and all-optical control of refractive index [52] have opened promising paths toward dynamic index modulation at optical frequencies. Photonic platforms such as lithium niobate modulators [53], transparent conducting oxides (TCOs) [54], and hybrid photonic–electronic materials [55] could serve as viable candidates for dynamic index modulation at the femtosecond scale.

**Conclusion.** In brief, we have proposed and demonstrated a new class of topological excitations—topological event wavepackets (TEWs)—in photonic spacetime crystals (STCs) with fully developed $\omega k$-gaps. These events emerge from spatiotemporal domain wall structures and exhibit strong confinement in both space and time, protected by a spacetime winding number. Our analysis of the $\omega k$-gapped band structure reveals that TEWs arise from the topological hybridization of four degenerate dissipative modes embedded within the gap. The degree of TEW localization is governed by the modulation amplitude and frequency, while their momentum-spectral width precisely tracks the size of the $\omega k$-gap—thus offering a direct measurement for bandgap engineering. To further exploit these event excitations, we introduced a spatiotemporal weaving strategy that periodically generates TEWs and effectively suppresses background noise amplification. Notably, we uncovered a non-monotonic dependence on the temporal weaving period, revealing the existence of an optimal regime for dynamical stabilization via topological control. Our results open a promising avenue for robust wave manipulation through topologically engineered $\omega k$-gaps.

**Acknowledgement**

Y. P. acknowledges the support of the NSFC (No. 2023X0201-417-03) and the fund of the ShanghaiTech University (Start-up funding).

Corresponding to: yiming.pan@shanghaitech.edu.cn

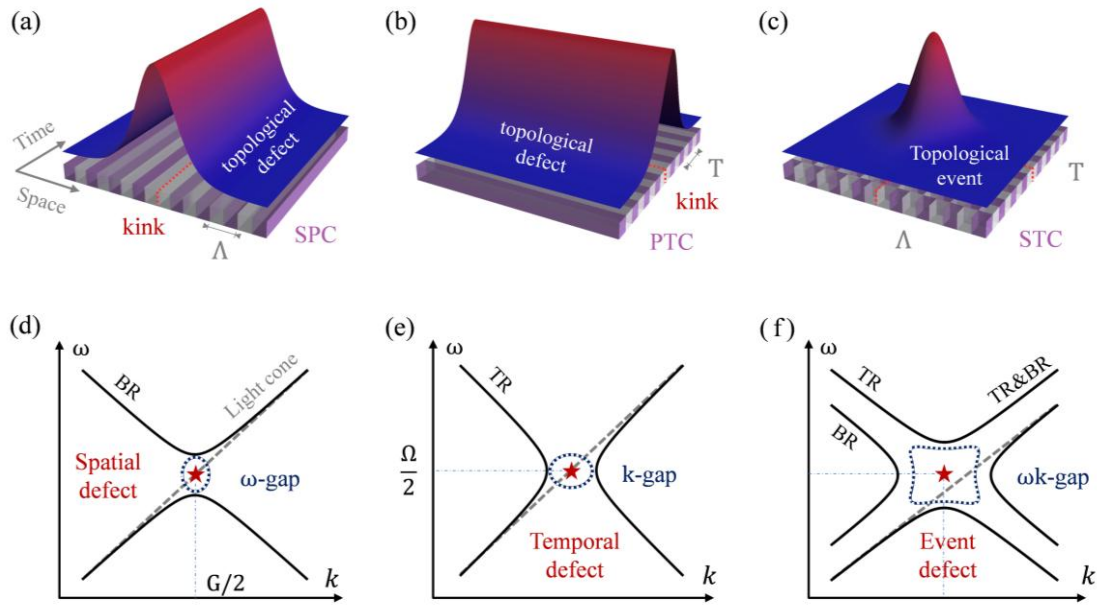

**Figure 1| Schematic comparison of topological defects and band structures of distinct photonic crystals.** The topological defects are: (a) Spatial topological edge state in photonic crystal (SPC); (b)Temporal topological edge state in photonic temporal crystal (PTC); (c) Topological event wavepacket in photonic spacetime crystal (STC). Corresponding band structures are: (d) Energy $\omega$-gapped band due to Bragg reflection; (e) Momentum k-gapped band due to temporal reflection; (f) Mixed $\omega k$-gapped band due to both Bragg and temporal reflections.

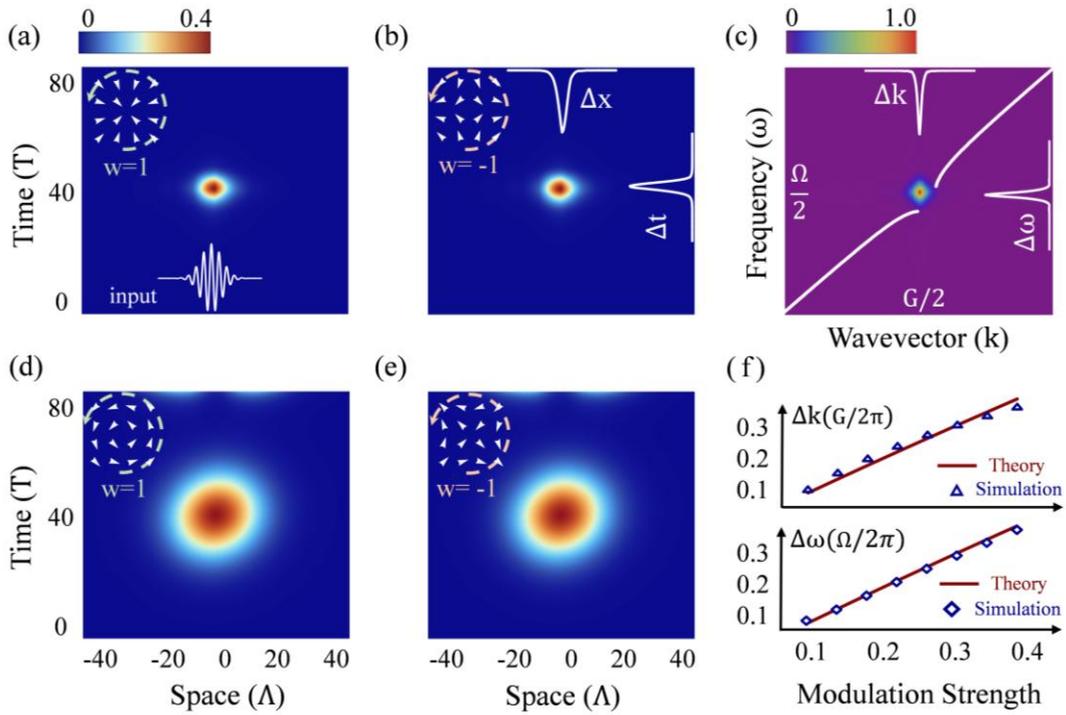

**Figure 2| Emergence and spectral localization of topological event wavepacket (TEW) in fully $\omega k$-gapped photonic spacetime crystals.** Define spacetime winding number of domain walls: w. (a, d) $w = 1$, (b, e) $w = -1$. (a, b) Topological event wavepacket is sharply localized in spacetime domain, with spatial and temporal widths $\Delta x$ and $\Delta t$. (c) Frequency-domain representation shows localization of the TEW inside the $\omega k$-gap, with spatial and temporal spectral widths $\Delta k$ and $\Delta \omega$. (d, e) Even when the modulation strengths $\kappa_1, \kappa_2$ are dependent on position and time, the topological event wavepacket remain confined. (f) Tuning of modulation strengths $\kappa_1$ and $\kappa_2$ controls the spectral width of the TEW. Numerical and theoretical results agree, confirming tunable confinement of the TEW in the frequency domain.

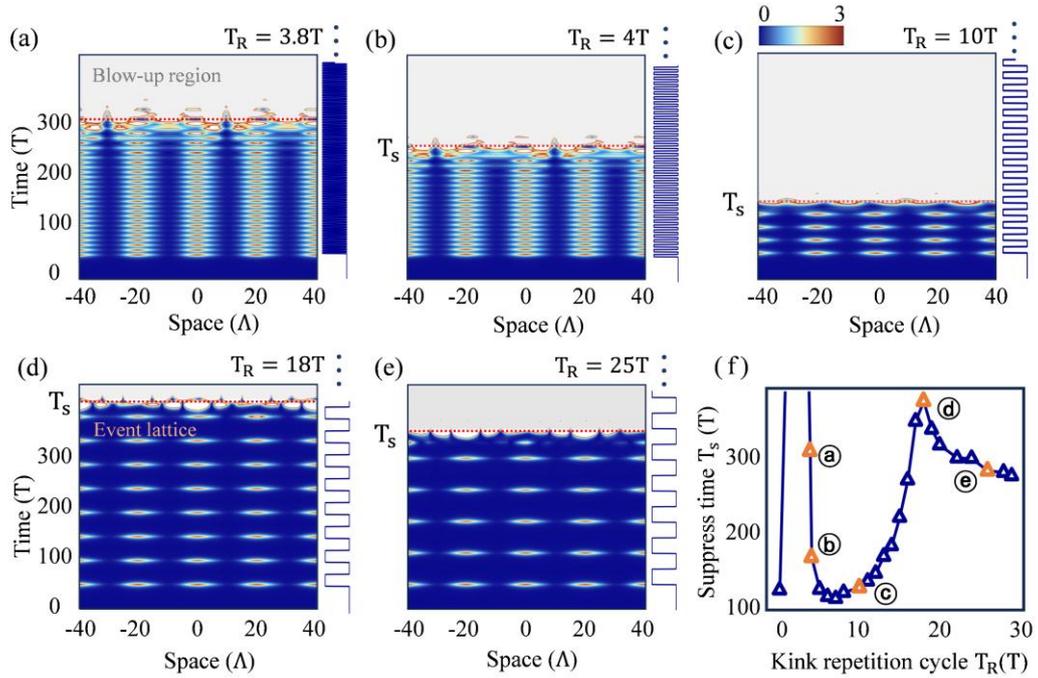

**Figure 3| Formation of topological event lattice through spatiotemporal weaving on suppression of $\omega k$ -gapped noise amplification.** (a-e) Spatial-temporal configurations of domain walls with different temporal repetition periods $T_R$, leading to periodic TEW arrays (event lattices). (f) Quantification of suppression time $T_s$ as a function of the repetition cycle $T_R$, showing how the lattice spacing controls the suppression of noise amplification within the $\omega k$-gap.